\documentclass[12pt]{article}
\usepackage{epsfig}
\begin{document}

\def\Omm{\Omega_m}
\def\Oml{\Omega_{\lambda}}
\def\Om0{\Omega_0}
\def\Omb{\Omega_b}
\def\Ombh{\Omega_b\thinspace h^2}

\input amssym.tex

\vspace*{5.00 cm}

\centerline{\bf\Large{The Dynamical Parameters of the Universe}}

\vspace*{2.00 cm}

\centerline{\bf Matts Roos and S. M. Harun-or-Rashid}

\vspace*{1.00 cm}

\centerline{Department of Physics, Division of High Energy
Physics,} \centerline{University of Helsinki, Finland}

\newpage

\centerline{\bf ABSTRACT} 

\vspace*{0.5cm}

\noindent The results of different analyses of the dynamical parameters 
of the Universe are converging towards agreement. Remaining
disagreements reflect systematic errors coming either from the
observations or from differences in the methods of analysis.
Compiling the most precise parameter values with our estimates of
such systematic errors added, we find the following best
values: the baryonic density parameter $\Ombh =0.019\pm 0.02$, the
density parameter of the matter component $\Omm =0.29\pm 0.06$,
the density parameter of the cosmological constant $\Oml = 0.71\pm
0.07$, the spectral index of scalar fluctuations $n_s =1.02 \pm
0.08$, the equation of state of the cosmological constant
$w_{\lambda} < -0.86$, and the deceleration parameter $q_0 = -0.56
\pm 0.04$. We do not modify the published best values of the
Hubble parameter $H_0 = 0.73\pm 0.07$ and the total density
parameter $\Om0\thinspace ^{+0.03}_{-0.02}$.
\\

\section{INTRODUCTION}
\parindent=5.0mm

Our information on the dynamical parameters of the Universe
describing the cosmic expansion comes from three different epochs.
The earliest is the Big Bang nucleosynthesis which occurred a
little over 2 minutes after the Big Bang, and which left its
imprint in the abundances of the light elements affecting the
baryonic density parameter $\Omega_b$. The discovery of
anisotropic temperature fluctuations in the cosmic microwave
background radiation at large angular scales (CMBR) by COBE-DMR
\cite{smot}, followed by small scale anisotropies measured in the
balloon flights BOOMERANG \cite{dber} and MAXIMA \cite{ba-ha}, by
the radio telescopes Cosmic Background Imager (CBI) \cite{pe-ma},
Very Small Array (VSA) \cite{scot} and Degree Angular Scale
Interferometer (DASI) \cite{halv} testify about the conditions in
the Universe at the time of last scattering, about
350\thinspace000 years after Big Bang. The analyses of the CMBR
power spectrum give information about every dynamical parameter,
in particular $\Omega_0$ and its components $\Omega_b,\ \Omega_m$
and $\Omega_{\lambda}$, and the spectral index $n_s$. For an
extensive review of CMBR detectors and results, see Bersanelli et
al. \cite{bersa}. Very recently, also the expected fluctuations in
the CMBR polarization anisotropies has been observed by DASI
\cite{kova}.

The third epoch is the time of matter structures: galaxy clusters,
galaxies and stars. Our view is limited to the redshifts we can
observe which correspond to times of a few Gyr after Big Bang.
This determines the Hubble constant, successfully done by the
Hubble Space Telescope (HST) \cite{free}, and the difference
$\Omega_{\lambda}-\Omega_m$ in the dramatic supernova Ia
observations by the High-z Supernova Search Team \cite{ries} and
the Supernova Cosmology Project \cite{perl}. The large scale structure
(LSS) and its power spectrum has been studied in the SSRS2 and
CfA2 galaxy surveys \cite{daco}, in the Las Campanas Redshift
Survey \cite{shec}, in the Abell-ACO cluster survey \cite{retz},
in the IRAS PSCz Survey \cite{saun} and in the 2dF Galaxy Redshift
Survey \cite{peac},\cite{coll}.
Various sets of CMBR data, supernova data and LSS data have been
analyzed jointly. We shall only refer to global analyses of the
now most recent CMBR power spectra and large scale distributions of
galaxies.

The list of other types of observations is really very long. To
mention some, there have been observations on the gas fraction in
X-ray clusters \cite{evrd}, on X-ray cluster evolution
\cite{ba-ek}, on the cluster mass function and the Ly$\alpha$
forest \cite{wein}, on gravitational lensing \cite{c-h-i}, on the
Sunyaev-Zel'dovich effect \cite{bi-ca}, on classical double radio
sources \cite{guer}, on galaxy peculiar velocities \cite{zeha}, on
the evolution of galaxies and star creation versus the evolution
of galaxy luminosity densities \cite{tota}.

In this review we shall cover briefly recent observations and
results for the dynami\-cal parameters $H_0,\ \Omega_b,\ \Omega_m,\
\Omega_{\lambda},\ \Omega_0,\ n_s,\ w_\lambda$ and $q_0$. In
Section 2 these parameters are defined in their theoretical
context, in Section 3 we turn to the Hubble parameter, and in
Section 4 to the baryonic density. The other parameters are
discussed in Sections 5 and 6, which are organized according to
observational method: supernov\ae\ in Section 5, CMBR and LSS in
Section 6. Section 7 summarizes our results.
\\

\section{THEORY}

The currently accepted paradigm describing our homogeneous and
isotropic Universe is based on the Robertson--Walker metric

$$\hbox{d}s^2=c^2\hbox{d}t^2-\hbox{d}l^2=c^2\hbox{d}t^2-R(t)^2\left({{\hbox{d}\sigma ^2}\over {1-k\sigma ^2}}
+\sigma ^2\hbox{d}\theta^2+\sigma ^2\hbox{sin}^2\theta\ \hbox{d}
\phi^2\right)\ \eqno(1)$$

\noindent and Einstein's covariant formula for the law of
gravitation,

$$G_{\mu\nu}={{8\pi G}\over {c^4}}T_{\mu\nu}\ .\eqno(2)$$

\noindent In Eq.~(1) d$s$ is the line element in four-dimensional
spacetime, $t$ is the time, $R(t)$ is the cosmic scale, $\sigma$
is the comoving distance as measured by an observer who follows
the expansion, $k$ is the curvature parameter, $c$ is the velocity
of light, and $\theta,\ \phi$ are comoving angular coordinates.
In Eq.~(2) $G_{\mu\nu}$ is the Einstein tensor describing the
curved geometry of spacetime, $T_{\mu\nu}$ is the energy-momentum
tensor, and $G$ is Newton's constant.

From these equations one derives Friedmann's equations which can
be put into the form

$${{\dot{R}^2 + kc^2}\over {R^2}}=
{{8\pi G}\over {3}}(\rho_m+\rho_{\lambda})\ ,\eqno(3)$$

$${{2\ddot{R}}\over {R}}+{{\dot{R}^2 + kc^2}\over {R^2}}=
-{{8\pi G}\over {c^2}}(p_m+p_\lambda)\ .\eqno(4)$$

\noindent Here $\rho$ are energy densities, the subscripts $m$ and
$\lambda$ refer to matter and cosmological constant (or dark
energy), respectively; $p_m$ and $p_{\lambda}$ are the
corresponding pressures of matter and dark energy, respectively.
Using the expression for the critical density today,

$$\rho_c={3\over{8\pi G}}H_0^2\ ,\eqno(5)$$

\noindent where $H_0$ is the Hubble parameter at the present time,
one can define density parameters for each energy component by

$$\Omega=\rho/\rho_c\ .\eqno(6)$$

\noindent The total density parameter is

$$\Omega_0=\Omega_m+\Omega_r+\Omega_{\lambda}\ .\eqno(7)$$

\noindent In what follows we shall ignore the very small radiation
density parameter $\Omega_r$. The matter density parameter
$\Omega_m$ can further be divided into a cold dark matter (CDM)
component $\Omega_{CDM}$, a baryonic component $\Omega_b$ and a
neutrino component $\Omega_{\nu}$.

The pressure of matter is certainly very small, otherwise one
would observe the galaxies having random motion similar to that of
molecules in a gas under pressure. Thus one can set $p_m=0$ in
Eq.~(4) to a good approximation. If the expansion is adiabatic so
that the pressure of dark energy can be written in the form

$$p_\lambda=w_\lambda \rho_\lambda c^2\ ,\eqno(8)$$

\noindent and if dark energy and matter do not transform into one
another, conservation of dark energy can be written

$$\dot{\rho_\lambda}+3H\rho_\lambda(1 + w_\lambda)=0\ .\eqno(9)$$

One further parameter is the deceleration parameter $q_0$, defined
by

$$q=-{{R\ddot R}\over {\dot{R}^2}}=-{{\ddot R}\over {RH^2}}\ .\eqno(10)$$

\noindent Eliminating $\ddot R$ between Eqs.~(4) and (10) one can
see that $q_0$ is not an independent parameter.

The curvature parameter $k$ in Eqs.~(1), (3) and (4) describes the
geometry of space: a spatially open universe is defined by $k=-1$,
a closed universe by $k=+1$ and a flat universe by $k=0$. The
curvature parameter is not an observable, but it is proportional
to $\Omega_0-1$, so if $\Omega_0$ is observed to be 1, the
Universe is spatially flat.\\


\section{THE HUBBLE PARAMETER}

From the definition of the Hubble parameter $H=\dot{R}/R$ one sees
that it has the dimension of inverse time. Thus a characteristic
time scale for the expansion of the Universe is the Hubble time

$$\tau_H\equiv H_0^{-1}=9.78h^{-1}\times 10^{9} \hbox{yr}.\eqno(11)$$

\noindent Here $h$ is the commonly used dimensionless quantity

$$h=H_0/(100\ \hbox{km\ s}^{-1}\ \hbox{Mpc}^{-1})\ .\eqno(12)$$

 The Hubble parameter also determines the size scale of
the observable Universe. In time $\tau_H$, radiation travelling
with the speed of light has reached the Hubble radius

$$r_H\equiv \tau_Hc = 3000 h^{-1}\hbox{Mpc}.\eqno(13)$$

\noindent Or, to put it differently, according to Hubble's
non--relativistic law,

$$z = H_0{{r}\over {c}}\ ,\eqno(14)$$

\noindent objects at this distance would be expected to attain the
speed of light which is an absolute limit in the theory of special
relativity. However, in special relativity the redshift $z$ is
infinite for objects at distance $r_H$ receding with the speed of
light and thus unphysical. Therefore no information can reach us
from farther away, all radiation is redshifted to infinite
wavelengths and no particle emitted within the Universe can exceed
this distance.

Our present knowledge of $H_0$ comes from the Hubble Space
Telescope (HST) Key Project \cite{free}.  The goal of this project
was to determine $H_0$ by a Cepheid calibration of a number of
independent, secondary distance indicators, including Type Ia
supernovae, the Tully-Fisher relation, the fundamental plane for
elliptical galaxies, surface brightness fluctuations, and type II
supernovae. Here we shall restrict the discussion to the best
absolute determinations of $H_0$, which are those from
supernov\ae\ of type Ia.

Visible bright supernova explosions are very brief events (one
month) and very rare, historical records show that in our Galaxy
they have occurred only every 300 years. The most recent one
occurred in 1987 (code name SN1987A), not exactly in our Galaxy
but in the nearby Large Magellanic Cloud (LMC). Since it now has
become possible to observe supernov\ae\ in very distant galaxies,
one does not have to wait 300 years for the next one.

The physical reason for this type of explosion (type SNII
supernova) is the accumulation of Fe--group elements at the core
of a massive red giant star of size 8--200 $M_{\odot}$ which
already has burned its hydrogen, helium and other light elements.
Another type of explosion (type SNIa supernova) occurs when a
degenerate dwarf star of CNO composition enters a stage of rapid
nuclear burning to Fe--group elements.

The SNIa is the brightest and most homogeneous class of
supernov\ae\ with hydrogen-poor spectra, their peak brightness can
serve as remarkably precise standard candles visible from very
far. Additional information is provided by the colour, the
spectrum, and an empirical correlation observed between the time
scale of the sharply rising light curve and the peak luminosity,
which is followed by a gradual decline. Although supernov\ae\ are
difficult to find, they can be used to determine $H_0$ out to
great distances, 500 Mpc or $z\approx 0.1$, and the internal
precision of the method is very high. At greater distances one can
still find supernov\ae , but Hubble's linear law (14) is then no
longer valid, the velocity starts to accelerate.

Supernov\ae ~of type II are fainter, and show a wider variation in
luminosity. Thus they are not standard candles, but the time
evolution of their expanding atmospheres provides an indirect
distance indicator, useful out to some 200 Mpc.

Two further methods to determine $H_0$ make use of correlations
between different galaxy properties. Spiral galaxies rotate, and
there the Tully-Fisher relation correlates total luminosity with
maximum rotation velocity. This is currently the most commonly
applied distance indicator, useful for measuring extragalactic
distances out to about 150 Mpc. Elliptical galaxies do not rotate,
they are found to occupy a "fundamental plane" in which an
effective radius is tightly correlated with the surface brightness
inside that radius and with the central velocity dispersion of the
stars. In principle this method could be applied out to $z\approx
1$, but in practice stellar evolution effects and the
non-linearity of Hubble's law limit the method to $z\lesssim 0.1$,
or about 400 Mpc.

The resolution of individual stars within galaxies clearly depends
on the distance to the galaxy.  This method, called surface
brightness fluctuations (SBF), is an indicator of relative
distances to elliptical galaxies and some types of spirals. The
internal precision of the method is very high, but it can be
applied only out to about 70 Mpc.

Observations from the HST combining all this methods \cite{free}
and independent SNIa observations from observatories on the ground
\cite{gibs} agree on a value

$$H_0=73\pm 2\pm 7\ \hbox{km\ s}^{-1}\ \hbox{Mpc}^{-1}.\eqno(15)$$

\noindent Note that the second error in Eq.~(15) which is
systematical, is much bigger than the statistical error. This
illustrates that there are many unknown effects which complicate
the determination of $H_0$, and which in the past have made all
determinations controversial. To give just one example, if there
is dust on the sight line to a supernova, its light would be
reddened and one would conclude that the recession velocity is
higher than it in reality is. There are other methods such as weak
lensing which do not suffer from this systematic error, but they
have not yet reached a precision superior to that in Eq.~(15).\\

\section{THE BARYONIC DENSITY}

The ratio of baryons to photons or the baryon abundance is defined
as

$$\eta\equiv{N_b\over N_{\gamma}}\simeq 2.75\times 10^{-8}\
\Ombh\,\eqno(16)$$

\noindent where $N_b$ is the number density of baryons and
$N_{\gamma}=4.11 \times 10^8\ \hbox{m}^{-3}$ is the number density
of photons. Thus the primordial abundances of baryonic matter in
the standard Big Bang nucleosynthesis scenario (BBN) is
proportional to $\Ombh$. Its value is obtained in direct
measurements of the abundances of the light elements $^4$He,
$^3$He, $^2$H or D, $^7$Li and indirectly from CMBR observations
and galaxy cluster observations.

If the observed abundances are indeed of cosmological origin, they
must not significantly be affected by later stellar processes. The
helium isotopes $^3$He and $^4$He cannot be destroyed easily but
they are continuously produced in stellar interiors. Some recent
helium is blown off from supernova progenitors, but that fraction
can be corrected for by observing the total abundance in hydrogen
clouds of different age, and extrapolating it to time zero. The
remainder is then primordial helium emanating from BBN. On the
other hand, the deuterium abundance can only decrease, it is
easily burned to $^3$He in later stellar events. The case of
$^7$Li is complicated because some fraction is due to later
galactic cosmic ray spallation products.

Among the light elements the $^4$He abundance is easiest to
observe, but also least sensitive to $\Ombh$, its
dependence is logarithmic, so that only very precise measurements
are relevant. The best "laboratories" for measuring the $^4$He
abundance are a class of low-luminosity dwarf galaxies called Blue
Compact Dwarf (BCD) galaxies, which undergo an intense burst of
star formation in a very compact region. The BCDs are among the
most metal-deficient gas-rich galaxies known. Since their gas has
not been processed during many generations of stars, it should
approximate well the pristine primordial gas.

Over the years the observations have yielded many conflicting
results, but the data are now progressing towards a common value
\cite{luri}, in particular by the work of Yu. I. Izotov and his
group. The analysis in their most recent paper \cite{izot}, based
on the two most metal-deficient BCDs known, gives the result

$$\Omega_b(^4\hbox{He})\thinspace h^2= 0.017\pm 0.005\ \ \ \ (2\sigma\ \hbox{CL})\ ,\eqno(17)$$

\noindent where the error is statistical only. Usually one quotes
the ratio $Y_p$ of mass in $^4$He to total mass in $^1$H and $^4$He,
which in this case is 0.2452 with a systematic error in the
positive direction estimated to be 2-4\%. Because of the
logarithmic dependence, this error translated to $\Ombh$
could be considerable, of the order of 100\% .

The $^3$He isotope can be seen in the Milky Way interstellar
medium and its abundance is a strong constraint on $\Ombh$.
The $^3$He abundance has been determined from 14 years of
data by Balser et al. \cite{bals}. More interestingly, Bania et
al. \cite{bani} combined Milky Way data with the helium abundance
in stars \cite{char} to find

$$\Omega_b(^3\hbox{He})\thinspace h^2 = 0.020^{+0.007}_{-0.003}\ \ \ (1\sigma\ \hbox{CL})\ .\eqno(18)$$

There are actually three different errors in their analysis, and
their quadratic sum gives the total error. The first error is from
the observed emission-line that includes the errors in the
Gaussian fits to the observed line parameters. The second error is
from the standard deviation of the observed continuum data and the
third error is the percent uncertainty of all models that have
been used in the analyses of reference \cite{bals}.

For a constraint on $\Ombh$ from $^7$Li, Coc et al. \cite{coca}
update the previous work of several groups. More importantly, they
include NACRE data \cite{nacre} in their compilation, and the
uncertainties are analysed in detail. There is some lack of
information about the neutron-induced reaction in the NACRE
compilation, but the main source of uncertainty for the lighter
neutron-induced reaction (e.g. ${\rm ^1H(n,\gamma)^2H}$ and ${\rm
^3He(n,p)^3H}$) is the neutron lifetime (for the present value see
the Review of Particle Physics \cite{grom}). However, there is no
new information about the heavier neutron-induced reaction (e.g.
${\rm ^7Li}$) or for ${\rm ^3He(d,p)^3He}$, but in this
compilation the Gaussian errors have been opted from the
polynomial fit of Nollett \& Burles \cite{noll}. We quote Coc et
al. \cite{coca} for

$$\Omega_b(^7\hbox{Li})\thinspace h^2= 0.015\pm 0.003\ \ \ \ (1\sigma\ \hbox{CL})\ .\eqno(19)$$

The strongest constraint on the baryonic density comes from the
primordial deuterium abundance. Deuterium is observed as a
Lyman-$\alpha$ feature in the absorption spectra of high-redshift
quasars. A recent analysis \cite{burl} gives

$$\Omega_b(^2\hbox{H})\thinspace h^2= 0.020\pm 0.001\ \ \ \ (1\sigma\ \hbox{CL})\ ,\eqno(20)$$

\noindent more precisely than any other determination. Some
systematic uncertainties remain in the calculations arising from
the reaction cross sections.

Very recently Chiappini et al. \cite{chia} have redefined the
production and destruction of ${\rm ^3He}$ in low and intermediate
mass stars. They also propose a new model for the time evolution
of deuterium in the Galaxy. Taken together, they conclude that
$\Ombh \gtrsim 0.017$, in good agreement with the values in
Eqs.~(18) and (20).

Let us now turn to the information from the cosmic microwave
background radiation and from large scale structures. There are
many analyses of joint CMBR data, in particular three large
compilations. Percival et al. \cite{perc} combine the data from
COBE-DMR \cite{smot} MAXIMA \cite{leea}, BOOMERANG \cite{nett},
DASI \cite{halv}, VSA \cite{scot} and CBI \cite{pe-ma} with the
2dFGRS LSS data \cite{coll}. Wang et al.\cite{wang} combine the
same $\hbox{CMBR}$ data (except VSA) with 20 earlier CMBR power
spectra, take their LSS power spectra from the IRAS PSCz survey
\cite{saun}, and include constraints from Lyman $\alpha$ forest
spectra \cite{crof} and from the Hubble parameter \cite{free}
quoted in Eq.~(15). Sievers et al. \cite{siev} also use the same
CMBR data as Percival et al.\cite{perc} (except VSA), combine them
with earlier LSS data, and use the HST Hubble parameter
\cite{free} quoted in Eq.~(15) and the supernova data referred to
in Section 5 as supplementary constraints. All these analyses are
maximum likelihood fits based on frequentist statistics, so the
use of the Bayesian term "prior" for constraint is a misnomer.

Assuming that the initial seed fluctuations were adiabatic,
Gaussian, and well described by power law spectra, the values of a
large number of parameters are obtained by fitting the observed
power spectrum. Here we shall only discuss results on
$\Ombh$ which is essentially measured by the relative magnitudes of
the first and second acoustic peaks in the CMBR power spectrum,
returning to this subject in more detail in Section 6.

The data used in the three compilations are overlapping but not
identical, and the central values show a spread over $\pm 0.0003$.
This we treat as a systematic error to the straight unweighted
average of the central values. Two compilations \cite{perc},
\cite{wang} consider models with and without a tensor component.
Since the fits are equally good in both cases we take their
difference, $\pm 0.0008$, to constitute another syste\-matic error.
We shall use this averaging prescription also in Section 6 to
obtain values of other parameters. All the analyses can then be
summarized by the value

$$\Omega_b(\hbox{CMBR})\thinspace h^2= 0.022\pm 0.002 \pm 0.001\ \ \ \
(1\sigma\ \hbox{CL})\ ,\eqno(21)$$

\noindent where the statistical error corresponds to references
\cite{perc}, \cite{wang}.

\begin{table}[h]
\begin{center}
\begin{tabular}{lllll}
\hline {\em Method}  & $\eta$ & $\Ombh$ & {\em Error} & {\em References} \\
\hline ${\rm ^4He}$ abundance  & ${\rm 4.7\ ^{+1.0}_{-0.8} \times
10^{-10}}$ & ${\rm 0.017\pm 0.005}$ & $2\sigma$ stat. only & \cite{izot}  \\
${\rm ^3He}$ abundance & ${\rm 5.4\ ^{+2.2}_{-1.2} \times
10^{-10}}$ & ${\rm 0.020\ ^{+0.007}_{-0.003}}$ & $1\sigma$ stat.
only &
\cite{bani}   \\
${\rm ^7Li}$ abundance  & ${\rm 5.0 \times 10^{-10}}$
& ${\rm 0.015 \pm 0.003}$ & $1\sigma$ stat. only  & \cite{coca} \\
$^2$H abundance  & ${\rm 5.6 \pm 0.5 \times 10^{-10}}$
& ${\rm 0.020 \pm 0.001}$ & $1\sigma$ stat.+syst. & \cite{burl}   \\
CMBR + 2dFGRS & ------ & ${\rm 0.022 \pm 0.002 \pm 0.001}$ & $1\sigma$
stat.+ syst.& \cite{perc}\cite{wang}\\
\hline
\end{tabular}
\end{center}
\caption {{\small The baryonic density parameter}} \label{table1}
\end{table}

In Table 1 we summarize the results from Eqs.~(17-21). From this
table one can conclude that all determinations are consistent with
the most precise one from deuterium \cite{burl}. A weighted mean
using the quoted errors yields $0.0194\pm 0.0008$ which is
dominated by deuterium. However, all light element abundance
determinations generally suffer from the potential for systematic
errors. As to CMBR, the statistical errors quoted in all
compilations have been obtained by marginalizing, so they are
certainly unrealistically small. We take a conservative approach
and add a systematic error of $\pm 0.002$ linearly to each of the
five data values before averaging. The weighted mean is then

$$\Ombh = 0.019\pm 0.002\ ,\eqno(22)$$

\noindent in excellent agreement with all the uncorrected input
values in Table 1.

One further source of $\Omega_b$ information is galaxy clusters
which are composed of baryonic and non-baryonic matter. The
baryonic matter takes the forms of hot gas emitting X-rays,
stellar mass observed in visual light, and perhaps invisible
baryonic dark matter of unknown composition. Let us denote the
respective fractions $f_{gas},\ f_{gal} $, and $ f_{bdm} $.  Then

$$f_{gas} + f_{gal} + f_{bdm} =
\Upsilon{\Omega_b\over\Omega_m}\ ,\eqno(21)$$

\noindent where $\Upsilon$ describes the possible local
enhancement or diminution of baryon matter density in a cluster
compared to the universal baryon density. This relation could in
principle be used to determine $\Omega_b$ when one knows
$\Omega_m$ (or vice versa), since $f_{gas}$ and $f_{gal} $ can be
measured, albeit with large scatter, while $ f_{bdm}$ can be
assumed negligible. Cluster formation simulations give information
on $\Upsilon$ \cite{eke},\cite{fre} to a precision of about 10\%.
However, the precision obtained for $\Ombh$ by adding
several 10\% errors in quadrature does not make this method
competitive.\\

\section{SUPERNOVA Ia CONSTRAINTS}

In Section 3 we already mentioned briefly the physics of
supernov\ae . The SN Ia observations by the High-z Supernova
Search Team (HSST) \cite{ries} and the Supernova Cosmology Project
(SCP) \cite{perl} are well enough known not to require a detailed
presentation here. The importance of these observations lies in
that they determine approximately the linear combination $\Oml -
\Omm$ which is orthogonal to $\Om0 = \Omm + \Oml$, see Figure 1.

\begin{figure*}
\setlength{\epsfxsize}{0.80\textwidth}
\centerline{\epsfbox{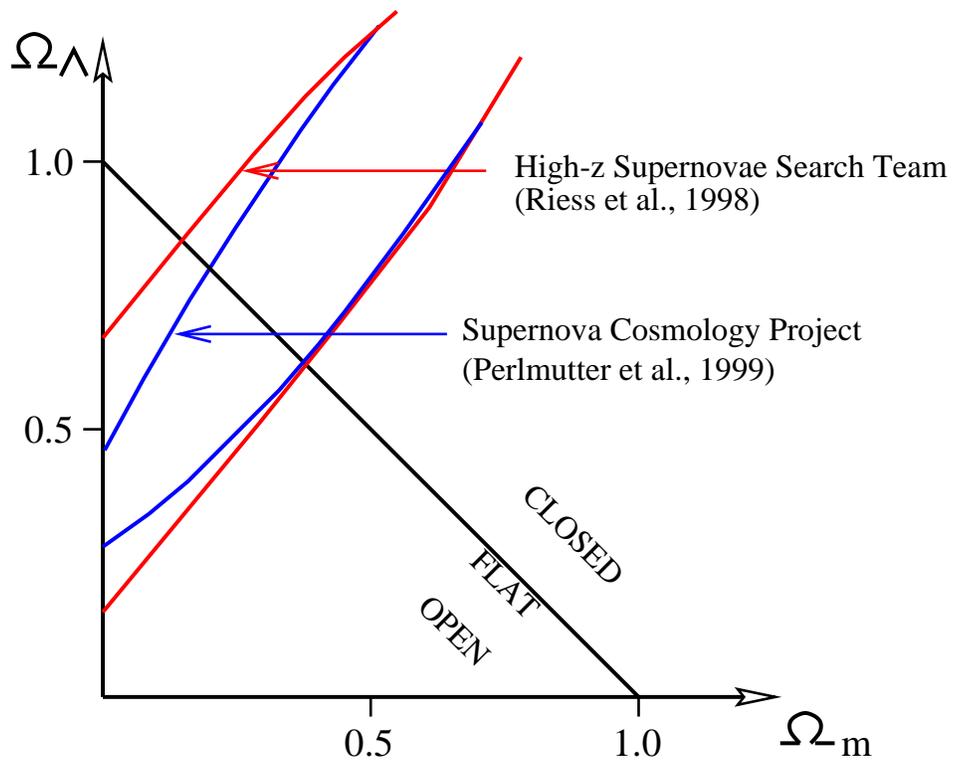}}

\caption{{\small The best fit confidence regions in $(\Omega_m -
\Omega_\lambda)$ plane in the analyses of the Supernova Cosmology
Project (blue curves) \cite{perl} and the High Redshift Supernova
Search Team (red curves) \cite{ries}. The diagonal line
corresponds to a flat cosmology. Above the flat line the Universe
is closed and below it is open.}}

\end{figure*}

HSST use two quite distinct methods of light-curve fitting to
determine the distance moduli of their 16 SNe Ia studied. Their
luminosity distances are used to place constraints on six
cosmological parameters: $h, \Omm, \Oml, q_0,$ and the dynamical
age of the Universe, $t_0$.  The MLCS method involves statistical
methods at a more refined level than the empirical template model.
The distance moduli are found from a $\chi^2$ analysis using an
empirical model containing four free parameters. The MLCS method
and the template method give moduli which differ by about
$1\sigma$. Once the distance moduli are known, the parameters h,
$\Omm,\ \Oml$ are determined by a maximum likelihood fit, and
finally the Hubble parameter is integrated out. (The results are
really independent of h.) One may perhaps be somewhat concerned
about the assumption that each modulus is normally
distributed. We have no reason to doubt that, but if the iterative
$\chi^2$ analysis has yielded systematically skewed pdf's, then
the maximum likelihood fit will amplify the skewness.

The authors state that "the dominant source of statistical
uncertainty is the extinction measurement". The main doubt raised
about the SN Ia observations is the risk that (part of) the
reddening of the SNe Ia could be caused by intervening dust rather
than by the cosmological expansion, as we already noted after
Eq.~(15). Among the possible systematic
errors investigated is also that associated with extinction. No
systematic error is found to be important here, but for such a
small sample of SNe Ia one can expect that the selection bias
might be the largest problem.

The authors do not express any view about which method should be
consi\-dered more reliable, thus noting that "we must consider the
difference between the cosmological constraints reached from the
two fitting methods to be a systematic uncertainty". We shall come
back to this question later. Here we would like to point out that
if one corrects for the unphysical region $\Omm < 0$ using the
method of Feldman \& Cousins \cite{feld}, the best value and the
confidence contours will be shifted slightly towards higher values
of $\Om0$. This shift will be more important for the MLCS method
than for the template method, because the former extends deeper
into the unphysical $\Omm$ region.

Let us now turn to SCP, which studied 42 SNe Ia. The MLCS method
described above is basically repeated, but modified in many details
for which we refer the reader to the source \cite{perl}. The distance
moduli are again found from a $\chi^2$ analysis using an empirical
model containing four free parameters, but this model is slightly
different from the HSST treatment. The parameters $\Omm$ and
$\Oml$ are then determined by a maximum likelihood fit to four
parameters, of which the parameters $\mathcal{M}_B$ (an absolute
magnitude) and $\alpha$ (the slope of the width-luminosity
relation) are just ancillary variables which are integrated out (h
does not enter at all). The likelihood contours in
$(\Omega_m - \Omega_\lambda)$ plane of both supernov\ae 
~projects (SCP and HSST) are shown in Figure-1. The authors
then correct the resulting
likelihood contours for the unphysical region $\Omm < 0$ using the
method of Feldman \& Cousins \cite{feld}. Since the number of SNe
Ia is here so much larger than in HSST, the effects of selection
and of possible systematic errors can be investigated more
thoroughly. SCP quotes a total possible systematic uncertainty to
$\Omm^{flat}$ and $\Oml^{flat}$ of 0.05.

If we compare the observations along the line defining a flat
Universe, SCP finds $\Oml - \Omm = 0.44 \pm 0.085 \pm 0.05$,
whereas HSST finds $\Oml - \Omm = 0.36 \pm 0.10$ for the MLCS
method and $\Oml - \Omm = 0.68 \pm 0.09$ for the template method.
Treating this difference as a systematic error of size $\pm 0.16$
the combined SCP result is $0.52 \pm 0.10 \pm 0.16$. SCP and HSST
then agree within their statistical errors -- how well they agree
cannot be established since they are not completely independent.
We choose to quote a combined HSST and SCP value

$$\Oml - \Omm = 0.5 \pm 0.1\ ,\eqno(22)$$

\noindent which excludes a flat de Sitter universe with $\Oml -
\Omm = 1$ by $5\sigma$, and excludes a flat Einstein -- de Sitter
universe with $\Oml - \Omm = -1$ by $10\sigma$.\\

\section{CMBR AND LSS CONSTRAINTS}

The most important source of information on the cosmological
parameters are the anisotro\-pies observed in the CMBR temperature
and polarization maps over the sky. The temperature angular power
spectrum has been measured and analyzed since 1992 \cite{smot},
whereas the polarization spectrum is very recent \cite{kova} and
has not yet been analyzed to obtain values for the dynamical
parameters. Given the temperature angular power spectrum, the
polarization spectrum is predicted with essentially no free
parameters. At the moment one can say that the temperature angular
power spectrum supports the current model of the Universe as
defined by the dynamical parameters obtained from the temperature
angular power spectrum.

Temperature fluctuations in the CMBR around a mean temperature in
a direction $\alpha$ on the sky can be analyzed in terms of the
autocorrelation function $C(\theta)$ which measures the average
product of temperatures in two directions separated by an angle
$\theta$,

$$C(\theta)=\left\langle{\delta T\over T}(\alpha){\delta T\over T}
(\alpha+\theta)\right\rangle\ .\eqno(23)$$

\noindent For small angles $(\theta)$ the temperature
autocorrelation function can be expressed as a sum of Legendre
polynomials $P_{\ell}(\theta)$ of order $\ell$, the wave number,
with coefficients or powers $a_{\ell}^2$,

$$C(\theta)={1\over
4\pi}\sum_{\ell=2}^{\infty}a_{\ell}^2(2\ell+1)P_{\ell}(\cos\theta)\
. \eqno(24)$$

\noindent All analyses start with the quadrupole mode $\ell=2$
because the $\ell=0$ monopole mode is just the mean temperature
over the observed part of the sky, and the $\ell=1$ mode is the
dipole anisotropy due to the motion of Earth relative to the CMBR.
In the analysis the powers $a_{\ell}^2$ are adjusted to give a
best fit of $C(\theta)$ to the observed temperature. The resulting
distribution of $a_{\ell}^2$ values versus $\ell$ is the power
spectrum of the fluctuations, see Figure 2. The higher the angular
resolution, the more terms of high $\ell$ must be included.

The exact form of the power spectrum is very dependent on
assumptions about the matter content of the Universe. It can be
parametrized by the vacuum density parameter $\Omega_k = 1 -
\Om0$, the total density parameter $\Om0$ with its components
$\Omm ,\ \Oml$, and the matter density parameter $\Omm$ withits
components $\Omb ,\ \Omega_{CDM},\ \Omega_{\nu}$. Further
parameters are the Hubble parameter {\it h}, the tilt of scalar
fluctuations $n_s$, the CMBR quadrupole normalization for scalar
fluctuations {\it Q}, the tilt of tensor fluctuations $n_t$, the
CMB quadrupole normalization for tensor fluctuations {\it r}, and
the optical depth parameter $\tau$. Among these parameters, really
only about six have an influence on the fit.

In Section 4 we already noted that the relative magnitudes of the
first and second acoustic peaks are sensitive to $\Omb$. The
position of the first acoustic peak in multipole $\ell$ - space is
sensitive to $\Om0$, which makes the CMBR information
complementary (and in $\Omm ,\ \Oml$ - space orthogonal) to the
supernova information. A decrease in $\Om0$ corresponds to a
decrease in curvature and a shift of the power spectrum towards
high multipoles. An increase in $\Oml$ (in flat space) and a
decrease in h (keeping $\Ombh$ fixed) both boost the peaks and
change their location in $\ell$ - space.

\begin{figure*}
\setlength{\epsfxsize}{0.80\textwidth}
\centerline{\epsfbox{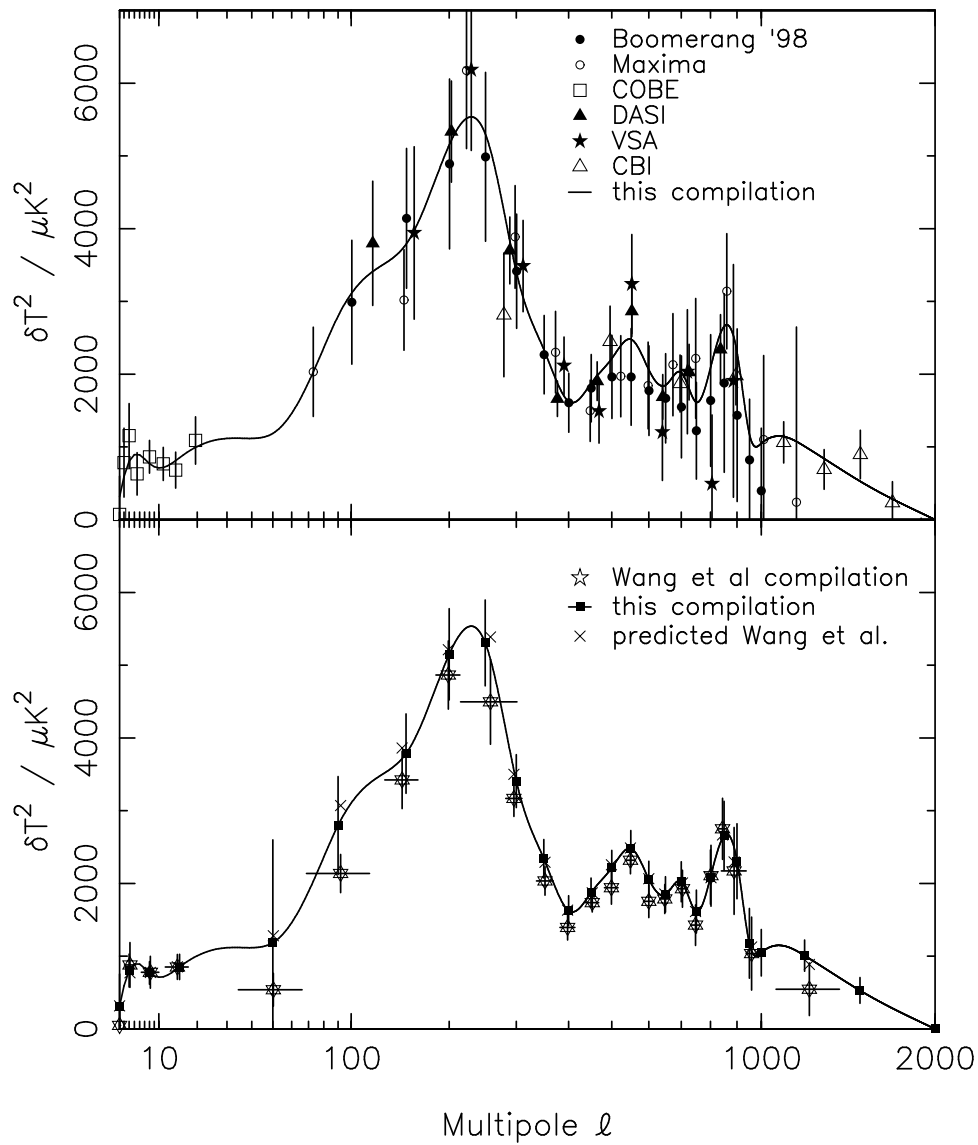}}

\caption{{\small Top panel: a compilation of recent CMB data
\cite{perc}. The solid line shows the result of a
maximum-likelihood fit to the power spectrum allowing for
calibration and beam uncertainty errors in addition to intrinsic
errors. Bottom panel: the solid line is as above, the solid
squares \cite{perc} and the crosses \cite{wang} give the points at
which the amplitude of the power spectrum was estimated. For
details, see reference \cite{perc}.}}

\end{figure*}

Let us now turn to the distribution of matter in the Universe
which can, to some approximation, be described by the
hydrodynamics of a viscous, non-static fluid. In such a medium
there naturally appear random fluctuations around the mean density
$\bar{\rho}(t)$, manifested by compressions in some regions and
rarefactions in other regions. An ordinary fluid is dominated by
the material pressure, but in the fluid of our Universe three
effects are competing: radiation pressure, gravitational
attraction and density dilution due to the Hubble flow. This makes
the physics different from ordinary hydrodynamics, regions of
overdensity are gravitationally amplified and may, if time
permits, grow into large inhomogenities, depleting adjacent
regions of underdensity.

Two complementary techniques are available for theoretical
modelling of galaxy formation and evolution:  numerical
simulations and semi-analytic modelling. The strategy in both
cases is to calculate how density perturbations emerging from the
Big Bang turn into visible galaxies. This requires following
through a number of processes: the growth of dark matter halos by
accretion and mergers, the dynamics of cooling gas, the
transformation of cold gas into stars, the spectrophotometric
evolution of the resulting stellar populations, the feedback from
star formation and evolution on the properties of prestellar gas,
and the build-up of large galaxies by mergers.

As in the case of the CMBR, an arbitrary pattern of fluctuations
can be mathematically described by an infinite sum of independent
waves, each with its characteristic wavelength $\lambda$ or
comoving wave number $k$ and its amplitude $\delta_k$. The sum can
be formally expressed as a  Fourier expansion for the density
contrast at comoving spatial coordinate {\bf r} and world time t,

$$\delta(\hbox{\bf r},t)\propto \sum\delta_k(t)\hbox{e}^{i{\bf k}\cdot{\bf
r}}\ ,\eqno(25)$$

\noindent where {\bf k} is the wave vector.

Analogously to Eq.~(23) a density fluctuation can be
expressed in terms of the dimensionless mass autocorrelation
function

$$\xi(r)=\langle\delta(\hbox{\bf r}_1)\delta(\hbox{\bf r+r}_1)\rangle\propto
\sum\langle|\delta_k(t)|^2\rangle e^{i{\bf k}\cdot{\bf r}}\
.\eqno(26)$$

\noindent which measures the correlation between the density
contrasts at two points {\bf r} and {\bf r}$_1$. The powers
$|\delta_k|^2$ define the power spectrum of the rms mass
fluctuations,

$$P(k)=\langle|\delta_k(t)|^2\rangle\ .\eqno(27) $$

\noindent Thus the autocorrelation function $\xi(r)$ is the
Fourier transform of the  power spectrum. This is similar to the
situation in the context of CMB anisotropies where the waves
represented temperature fluctuations on the surface of the
surrounding sky, and the powers $a_{\ell}^2$ were coefficients in
the Legendre polynomial expansion Eq.~(24).

With the lack of more accurate knowledge of the power spectrum one
assumes for simplicity that it is specified by a power law

$$P(k)\propto k^{n_s}\ ,\eqno(28)$$

\noindent where $n_s$ is the spectral index of scalar
fluctuations. Primordial gravitational fluctuations are expected
to have an equal amplitude on all scales. Inflationary models also
predict that the power spectrum of matter fluctuations is almost
scale-invariant as the fluctuations cross the Hubble radius. This
is the Harrison--Zel'dovich spectrum, for which $n_s=1$ ($n_s$ = 0
would correspond to white noise).

Since fluctuations in the matter distribution has the same
primordial cause as CMBR fluctuations, we can get some general
information from CMBR. There, increasing $n_s$ will raise the
angular spectrum at large values of $\ell$ with respect to low
$\ell$. Support for $\ell\approx 1.0$ come from all the available
analyses: combining the results of references \cite{perc},
\cite{wang}, \cite{siev} by the averaging prescription in Section
4, we find

$$n_s = 1.02 \pm 0.06 \pm 0.05\ .\eqno(29)$$

Phenomenological models of density fluctuations can be specified
by the amplitudes $\delta_k$ of the autocorrelation function
$\xi(r)$. In particular, if the fluctuations are Gaussian, they
are completely specified by the power spectrum $P(k)$. The models
can then be compared to the real distribution of galaxies and
galaxy clusters, and the phenomenological parameters determined.

As we noted in Section 4, there are several joint compilations of
CMBR power spectra and LSS power spectra of which we are
interested in the three largest ones \cite{perc}, \cite{wang},
\cite{siev}. Combining their results for $\Omm$ by the averaging
prescription in Section 4, we find

$$\Omm = 0.29 \pm 0.05 \pm 0.04\ .\eqno(30)$$

If the Universe is spatially flat so that $\Om0=1$, this gives
immediately the value $\Oml=0.71$ with slightly better precision
than above. To check this assumption we can quote reference
\cite{siev} from their Table 5 where they use all data,

$$\Om0 = 1.00 \pm\ ^{0.03}_{0.02}\ .\eqno(31)$$

\noindent Note, however, that this result has been obtained by marginalizing
over all other parameters, thus its small statistical errors are conditional
on $n_s,\ \Omm, \ \Omb$ being anything, and we have no
prescription for estimating a systematic error.

A value for $\Oml$ can be found by adding $\Oml-\Omm$ in Eq.~(22)
to $\Omm$, thus $\Oml = 0.79 \pm 0.12$. A better route appears to
be to combine Eqs.~(30) and (31) to give

$$\Oml = 0.71 \pm 0.07\ .\eqno(32)$$

Still a third route is to add $\Om0$ and $\Oml-\Omm$, or to
subtract them, respectively. Then one obtains

$$\Omm = 0.25 \pm 0.05\ ,\ \ \ \ \Oml = 0.75 \pm 0.05\ .$$

\noindent The routes making use of $\Oml-\Omm$ from Eq.~(22) are,
however, making multiple use of the supernova information, so we
discard them.

Before ending this Section, we can quote values also for
$w_\lambda$ and $q_0$. The notation here implies that $w_\lambda$
is taken as the equation of state of a quintessence component, so that
its value could be $w_\lambda > -1$. The equation of state of a
cosmomological constant component is of course $w_\lambda = -1$.
In a flat universe $w_\lambda$ is completely correlated to $\Oml$
and therefore also to $\Omm$.

We choose to quote the analysis by Bean and Melchiorri \cite{bean}
who combine CMBR power spectra from COBE-DMR \cite{smot}, MAXIMA
\cite{leea}, BOOMERANG \cite{nett}, DASI \cite{halv}, the
supernova data from HSST \cite{ries} and SCP \cite{perl}, the HST
Hubble constant \cite{free} quoted in Eq.~(15), the baryonic
density parameter $\Omb\ h^2 = 0.020 \pm 0.005$ and some LSS
information from local cluster abundances. They then obtain
likelihood contours in the $w_\lambda\ ,\Omm$ space from which
they quote the $1\sigma$ bound $w_\lambda < -0.85$. If we permit
ourselves to restrict their confidence range further by using our
value $\Omm = 0.29 \pm 0.06$ from Eq.~(30), the result is changed
only slightly to

$$w_\lambda < -0.86\ ,\ \ \ \ (1\sigma \ \hbox{CL}) \eqno(33).$$

Finally, the deceleration parameter is not an independent
quantity, it can be calculated from

$$q_0 = \hbox{${{1}\over {2}}$}\Omm - \Oml = \hbox{${{3}\over {2}}$}\Omm -
\Om0 = -0.56 \pm 0.04\ .\eqno(34)$$

\noindent The error is so small because the $\Omm$ and the $\Oml$
errors are completely anticorrelated. Note that the negative value
implies that the expansion of the Universe is accelerating.

\begin{table}[h]
\begin{center}
\begin{tabular}{lll}
\hline {\em Parameters}  & Values & {\em References} \\
\hline $H_0$ & ${\rm 73 \pm 7}$ & \cite{free}  \\
$\Ombh$ & ${\rm 0.019 \pm 0.002}$  & our compilation  \\
$\Omega_m$ & ${\rm 0.29 \pm 0.06}$  & our compilation  \\
$\Omega_{\lambda}$ & ${\rm 0.71 \pm 0.07}$ & our compilation  \\
$\Omega_0$ & ${\rm 1.0\ ^{+0.03}_{-0.02}}$  & \cite{siev}  \\
$n_s$ & ${\rm 1.02 \pm 0.08}$  & our compilation  \\
$w_{\lambda}$ & {\rm $<$ - 0.86} &  our compilation \\
$q_0$ & ${\rm -\ 0.56 \pm 0.04}$  & our compilation  \\
\hline
\end{tabular}
\end{center}
\caption {{\small Best values of the dynamical parameters. The errors
include $1\sigma$ statistical errors and our estimates of
systematic errors, except for $\Omega_0$ which is statistical
only. The Hubble constant $H_0$ is given in units of ${\rm km \
s^{-1}Mpc^{-1}}$ }} \label{table2}
\end{table}

\section{SUMMARY}

Information on the dynamical parameters of the Universe are coming
from the Big Bang nucleosynthesis, from the fluctuations in the
temperature and polarization of the cosmic microwave background
radiation, from the large scale structures of galaxies, from
supernova observations and from many other cosmological effects
that may not yet be of interesting precision. The results of
different analyses are now converging towards agreement when in
the past disagreements of the order of 100\%  have been known.

In this review we have taken the attitude that remaining
disagreements reflect systematic errors coming either from the
observations or from differences in the methods of analysis. We
have then compiled the most precise parameter values, combined
them and added our estimates of such systematic errors. This we
have done for the baryonic density parameter $\Ombh$, the density
parameter of the matter component $\Omm$, the density parameter of
the cosmological constant $\Oml$, the spectral index of scalar
fluctuations $n_s$, the equation of state of the cosmological
constant $w_{\lambda}$, and the deceleration parameter $q_0$. In
addition we quote the best values of the Hubble parameter $H_0$
and the total density parameter $\Om0$ from other sources.  In
Table 2 we summarize our results.

The conclusion is not new: that the Universe is spatially flat,
that some 25\%  of gravitating matter is dark and unknown, and
that some 70\%  of the total energy content is dark, possibly in
the form of a cosmological constant. \\

\noindent{\bf \it Acknowledgements:} S. M. H. is indebted to the Magnus 
Ehrnrooth Foundation for support.



\end{document}